\documentstyle[11pt]{article}
\newcount\Mac  \Mac=0 
\newcommand{\phit}{\tilde{\phi}}
\newcommand{\xt}{\tilde{x}}
\newcommand{\Vt}{\tilde{V}}

\newcommand{\lx}{\lambda}
\newcommand{\Lx}{\Lambda}

\newcommand{\kx}{\kappa}

\newcommand{\Gammak}{\Gamma_k}

\newcommand{\be}{\begin{equation}}
\newcommand{\ee}{\end{equation}}
\newcommand{\een}{\end{subequations}}
\newcommand{\ben}{\begin{subequations}}
\newcommand{\beq}{\begin{eqnarray}}
\newcommand{\eeq}{\end{eqnarray}}

\def \lta {\mathrel{\vcenter
     {\hbox{$<$}\nointerlineskip\hbox{$\sim$}}}}
\def \gta {\mathrel{\vcenter
     {\hbox{$>$}\nointerlineskip\hbox{$\sim$}}}}

\def\Red{}
\def\Black{}
\def\Blue{}

\newcommand{\lascia}[1]{}
\def\puttag(#1,#2)#3{\put(#1,#2){\makebox(0,0){\rm\Blue #3\Black}}}

\def\circa#1{\,\raise.3ex\hbox{$#1$\kern-.75em\lower1ex\hbox{$\sim$}}\,}

\newcommand{\riga}[1]{\noalign{\hbox{\parbox{\textwidth}{#1}}}\nonumber}

\def\putps(#1,#2)(#3,#4)#5#6{\ifnum\Mac=1 \put(#1,#2){\special{picture #5}}
\else  \put(#3,#4){\includegraphics{#6}} \fi}

\def\Op{{\cal W}}
\def\One{\hbox{1\kern-.24em I}}
\newcommand{\phibounce}{\phi_{\rm b}}

\makeatletter
\def\art{\@ifnextchar[{\eart}{\oart}}
\def\eart[#1]#2#3#4#5#6{{\rm #2}, {\e, #3 \bf #4} {\rm (#6) #5} ({\em #1})}
\def\hepart[#1]#2{{\rm #2, \em#1}}
\newcommand{\oart}[5]{{\rm #1}, {\em #2 \bf #3} {\rm (#5) #4}}
\newcounter{alphaequation}[equation]
\def\thealphaequation{\theequation\hbox to
0.6em{\hfil\alph{alphaequation}\hfil}}
\def\eqnsystem#1{
\def\@eqnnum{{\rm (\thealphaequation)}}
\def\@@eqncr{\let\@tempa\relax \ifcase\@eqcnt \def\@tempa{& & &} \or
  \def\@tempa{& &}\or \def\@tempa{&}\fi\@tempa
  \if@eqnsw\@eqnnum\refstepcounter{alphaequation}\fi
\global\@eqnswtrue\global\@eqcnt=0\cr}
\refstepcounter{equation} \let\@currentlabel\theequation \def\@tempb{#1}
\ifx\@tempb\empty\else\label{#1}\fi
\refstepcounter{alphaequation}
\let\@currentlabel\thealphaequation
\global\@eqnswtrue\global\@eqcnt=0 \tabskip\@centering\let\\=\@eqncr
$$\halign to \displaywidth\bgroup \@eqnsel\hskip\@centering
$\displaystyle\tabskip\z@{##}$&\global\@eqcnt\@ne
\hskip2\arraycolsep\hfil${##}$\hfil& \global\@eqcnt\tw@\hskip2\arraycolsep
$\displaystyle\tabskip\z@{##}$\hfil
\tabskip\@centering&\llap{##}\tabskip\z@\cr}

\def\endeqnsystem{\@@eqncr\egroup$$\global\@ignoretrue} \makeatother

\begin{document}
\begin{quote}
{\em April 1999}\hfill {\bf SNS-PH/99-4}\\
hep-ph/9904246
\hfill{\bf IFUP-TH/99-15}
\end{quote}
\vspace{2cm}
\begin{center}
{\LARGE\bf\Red Testing Nucleation Theory in Two Dimensions}\\
[2cm]
\Black\large
{\bf Alessandro Strumia}\\[0.3cm]
\normalsize\em 
Dipartimento di Fisica, Universit\`a di Pisa and\\
INFN, Sezione di Pisa, I-56127 Pisa, Italia\\[0.3cm]
\large {\rm and}\\[3mm] {\bf Nikolaos Tetradis}\\[3mm]
\normalsize\em  Scuola Normale Superiore,\\
Piazza dei Cavalieri 7, I-56126 Pisa, Italia\\[15mm]
\Blue\large\bf Abstract
\end{center}
\begin{quote}\large\indent
We calculate bubble-nucleation rates for (2+1)-dimensional
scalar theories at high temperature. Our approach is based
on the notion of a real coarse-grained potential. 
The region of applicability of our method is determined through internal
consistency criteria.
We compare our results with data from lattice simulations.
Good agreement is observed when
the renormalized action of the simulated theory is known.

\end{quote}\Black

\thispagestyle{empty}\newpage\setcounter{page}{1}

\setcounter{equation}{0}
\renewcommand{\theequation}{\thesection.\arabic{equation}}

\section{Introduction}

\subsection{The standard formalism}
The standard formalism for the 
calculation of bubble-nucleation rates during first-order
phase transitions in field theories at non-zero temperature 
was introduced in refs.~\cite{coleman}--\cite{linde}.
It consists of an implementation of 
Langer's theory of homogeneous
nucleation~\cite{langer} (for a review see ref.~\cite{review})
within the field-theoretical context. 
The nucleation rate $I$
gives the probability per unit time and volume to nucleate a certain
region of the stable phase (the true vacuum) within the metastable 
phase (the false vacuum).
Its calculation relies on a semiclassical approximation
around a dominant saddle-point, which is identified with 
the critical bubble. 
This is a static configuration
(usually assumed to be spherically symmetric) within the metastable phase 
whose interior consists of the stable phase.
It has a certain radius that can be determined from the
parameters of the underlying theory. Bubbles slightly larger
than the critical one expand rapidly, thus converting the 
metastable phase into the stable one. 
The nucleation rate is exponentially suppressed by the action
(the free energy rescaled by the temperature) of the critical bubble.
Possible deformations of the critical  bubble
generate a pre-exponential factor.
The leading contribution to this factor  
has the form of a ratio of fluctuation determinants and corresponds to the
first-order correction to the semiclassical result in a systematic 
expansion around the saddle point.

For a $(d+1)$-dimensional theory of a real scalar field 
at temperature
$T$, in the limit that thermal fluctuations dominate over
quantum fluctuations, the bubble-nucleation rate
is given by~\cite{colcal}--\cite{linde}
\be
I=\frac{E_0}{2\pi}
\left(\frac{S}{2\pi }\right)^{d/2}\left|
\frac{\det'[\delta^2 \Gamma/\delta\phi^2]_{\phi=\phibounce}}
{\det[\delta^2 \Gamma/\delta\phi^2]_{\phi=0}}\right|^{-1/2}
\exp\left(-S\right). 
\label{rate0} \ee
Here $\Gamma$ is the free energy of the system for a given configuration of the
field $\phi$.
The rescaled
free energy of the critical bubble is $S=\Gamma_b/T
=\left[\Gamma\left(\phibounce(r)\right)-\Gamma(0)\right]/T$,
where $\phibounce(r)$ is the spherically-symmetric
bubble configuration and $\phi = 0$ corresponds to the false vacuum. 
The prime in the fluctuation determinant around
the bubble denotes that the $d$ zero eigenvalues 
of the operator $[\delta^2 \Gamma/\delta\phi^2]_{\phi=\phibounce}$,
corresponding to displacements of the bubble,
have been removed. 
Their contribution generates the factor 
$\left(S/2\pi \right)^{d/2}$ and the volume factor
that is absorbed in the definition of $I$ (nucleation rate per unit volume). 
The quantity $E_0$ is the square root of
the absolute value of the unique negative eigenvalue.

In field theory, the free energy 
density $\Gamma$ of a system for homogeneous configurations 
is usually identified with
the temperature-dependent effective potential. This is evaluated 
through some perturbative scheme, such as the loop expansion.
The profile and the free energy of the critical bubble are determined
through the potential. This approach, however,
faces fundamental difficulties:
For example, the effective potential, being the Legendre transform of the
generating functional for the connected Green functions, 
is a convex function of the field. Consequently, it does not 
seem to be the appropriate quantity for the study of tunnelling.
Also,
the fluctuation determinants in the expression for the nucleation
rate have a form completely analogous to the one-loop correction to
the potential. The question of double-counting the effect of
fluctuations (in the potential and the prefactor)
must be properly addressed. 
A closely related issue 
concerns the ultraviolet divergences that are inherent in
the calculation of the fluctuation determinants 
in the prefactor. An appropriate regularization scheme must be
employed in order to control them~\cite{previous}. 
Moreover, this scheme must be consistent with the one employed
for the absorption of the divergences appearing in the 
calculation of the potential.

\subsection{Coarse-graining}
In refs.~\cite{bubble}--\cite{third}
it was shown that all the above issues can be resolved
through the implemention of the notion of coarse graining in the 
formalism. The appropriate quantity for the description of the
physical system is
the effective average action $\Gamma_k$~\cite{averact}, which 
is the generalization in the continuum   
of the blockspin action 
of Kadanoff~\cite{kadanoff}. It can be interpreted as a coarse-grained 
free energy at a given scale $k$. 
Fluctuations with 
characteristic momenta $q^2 \gta k^2$ are integrated out
and their effect is incorporated in 
$\Gamma_k$. 
In the limit $k \to 0$,
$\Gamma_k$ becomes equal to the effective action.
The $k$ dependence 
of $\Gamma_k$ is described by an exact flow equation~\cite{exact},
typical of the Wilson approach to the renormalization group~\cite{wilson}. 
This flow equation can be translated into evolution equations
for the functions appearing in a derivative expansion of
the action~\cite{indices,morris}. 
Usually, one 
considers only the effective average potential
$U_k$ and a standard kinetic term, and neglects higher derivative
terms in the action. We shall employ this approximation
in this paper also. 
The bare theory is defined
at some high scale $\Lx$ that can be identified with the ultraviolet 
cutoff. 
At scales $k$ below the temperature $T$,
a $(d+1)$-dimensional theory at non-zero temperature
can be described in terms of an effective  
$d$-dimensional action at zero temperature~\cite{trans}.

In ref.~\cite{first} we considered a (3+1)-dimensional
theory of a real scalar field at non-zero temperature,
defined through its action
$\Gamma_{k_0}$ 
at a scale $k_0$ below the temperature, so that 
the theory has an effective three-dimensional description.
The form of the potential $U_{k_0}$
results from the bare potential 
$U_{\Lx}$ after the integration of (quantum and thermal)
fluctuations between the scales
$\Lx$ and $k_0$. 
We computed the form of the $U_k$ at scales $k\leq k_0$ by
integrating an evolution equation derived from the exact flow 
equation for $\Gamma_k$.  
$U_k$ is non-convex for non-zero $k$, and 
approaches convexity only in the limit $k\to 0$.
The nucleation rate must be computed for $k$ larger than the scale $k_f$
at which the functional integral in the definition of $U_k$ starts 
receiving contributions from field configurations that interpolate between
the two minima. This happens when $-k^2$ becomes approximately equal to 
the negative curvature at the top
of the barrier~\cite{convex}. 
For $k \gta k_f$ the typical length scale of a thick-wall critical
bubble is $\gta  1/k$. 
We performed the calculation of the nucleation rate for a range of scales
above and near $k_f$, for which $U_k$ is non-convex. 
In our approach 
the pre-exponential factor is well-defined and finite,
as an ultraviolet cutoff of order $k$ is implemented in the calculation of
the fluctuation determinants, so that fluctuations
with characteristic momenta $q^2 \gta k^2$ are not included. 
This is a natural consequence of the fact 
that all fluctuations with typical momenta above $k$ are
already incorporated in the form of $U_k$.
This modification also resolves naturally the problem of double-counting 
the effect of the fluctuations.

We found that the saddle-point configuration 
has an action $S_k$ with a significant $k$ dependence. 
For strongly first-order phase transitions, the nucleation
rate $I = A_k \exp(-S_k)$
is dominated by the exponential suppression.
The main role of the prefactor $A_k$, which is also $k$ dependent,  
is to remove the scale dependence from the total nucleation rate.
Thus, this physical quantity is independent of the scale $k$ that we 
introduced as a calculational tool.
The implication of our results 
is that the critical bubble should not be identified
just with the saddle point of the semiclassical approximation.
It is the combination of 
the saddle point and its possible deformations
in the thermal bath (accounted for by the fluctuation 
determinant in the prefactor) that has physical meaning. 
We also found that, 
for progressively more weakly first-order phase transitions,
the difference between 
$S_k$ and $\ln ( A_k/k^4_f )$ diminishes.
This indicates that the effects of fluctuations
become more enhanced. 
At the same time, a significant $k$ dependence of the
predicted nucleation rate develops. 
The reason for the above deficiency is clear. 
When the nucleation rate is roughly equal to or smaller 
than the contribution from the prefactor, the effect of 
the next order in the expansion around the saddle point is 
important and can no longer be neglected.
This indicates that there is a limit to the validity of Langer's picture 
of homogeneous nucleation. The region of validity of this picture
was investigated in detail in ref.~\cite{second}. First-order phase transitions
in two-scalar models were studied in ref.~\cite{third} and the consistency of
the approach summarized above was reconfirmed.
Moreover, the applicability of homogeneous nucleation theory to
radiatively-induced first-order phase transitions was tested. 
It was found that the expansion around the semiclassical saddle point
is not convergent for such phase transitions. This indicates that
estimates of bubble-nucleation rates for the electroweak phase transition
that are based only on the saddle-point action may be very misleading. 

\subsection{Plan of the paper}
In this paper we present an application of our formalism to 
(2+1)-dimensional theories at non-zero temperature. Our investigation provides
a new test of several points of our approach that depend strongly on the
dimensionality, such as the form of the 
evolution equation of the potential, the nature of the
ultraviolet divergences of the fluctuation determinants, 
and the $k$ dependence of
the saddle-point action and prefactor. The
complementarity between the $k$ dependence of $S_k$ and $A_k$ is a 
crucial requirement for the nucleation rate $I$ to be $k$ independent.
A strong motivation for this study stems from the existence of 
lattice simulations of nucleation for (1+1) and (2+1)-dimensional systems
\cite{onepone,twopone}. In particular, we shall compare our predictions 
for the nucleation rate with the lattice results of ref.~\cite{twopone}.

In the next section we summarize the basic steps of our method
and derive the necessary expressions for the calculation of the 
nucleation rate. In section 3 we present sample calculations
in two dimensions. In section 4 we apply our formalism to theories
that have been studied through lattice simulations and compare with
the lattice results. Our conclusions are presented in section 5.

\setcounter{equation}{0}
\renewcommand{\theequation}{\thesection.\arabic{equation}}

\section{The calculation of bubble-nucleation rates}
\subsection{Evolution equation for the potential}

We consider a model of a real scalar field $\phi$ in 2+1 dimensions.
The 
effective average action $\Gamma_k(\phi)$~\cite{averact}
is obtained by 
adding an infrared cutoff term to the bare action,
so that contributions from modes with 
characteristic momenta $q^2 \lta k^2$ are not taken into account.
We use
the simplest choice of a mass-like cutoff term $\sim k^2 \phi^2$,
for which the 
perturbative inverse propagator for massless fields is
$P_k(q) \sim q^2 + k^2$. This choice makes
the calculation of the fluctuation determinants in the pre-exponential factor 
of the nucleation rate technically feasible.
Subsequently,
the generating functional for the connected Green functions is defined,
from which the
generating functional for the 1PI Green functions can be obtained through
a Legendre transformation.
The presence of the modified propagator in the above definitions
results in the effective integration of the 
fluctuations with $q^2 \gta k^2$ only. Finally, the 
effective average action is 
obtained by removing the infrared cutoff from the
generating functional for the 1PI Green functions. 

The effective average action $\Gamma_k$ obeys an exact 
flow equation,
which describes its response to variations of the infrared cutoff $k$
~\cite{exact}. This can be turned into evolution equations for the
functions appearing in a derivative expansion of $\Gamma_k$~\cite{indices}.
In this work we use an approximation which neglects higher derivative 
terms in the action and approximates it by 
\be
\Gammak = 
\int d^2x \left\{ 
\frac{1}{2} \partial^{\mu} \phi~\partial_{\mu} \phi~  
+ U_k(\phi) \right\}.
\label{twoeleven} \ee
The above action describes the effective two-dimensional theory that 
results from the dimensional reduction of a high-temperature 
(2+1)-dimensional theory at scales below the temperature. 
The temperature has been absorbed in a redefinition of the fields and
their potential, so that these have dimensions appropriate
for an effective two-dimensional theory. The correspondence 
between the quantities we use and the ones of the (2+1)-dimensional
theory is given by 
\beq
\phi =\frac{\phi_{2+1}}{\sqrt{T}},\qquad
U(\phi)=\frac{U_{2+1}\left( \phi_{2+1},T \right)}{T}.
\label{fivethree} \eeq
In this way, the temperature does not appear explicitly in
our expressions. This has the additional advantage of 
permitting the straightforward application of our results
to the problem of quantum tunnelling in a two-dimensional 
theory at zero temperature~\cite{first}. 

The evolution equation for the potential
can be written in the form~\cite{exact,indices,second}
\be 
\frac{\partial}{\partial k^2} \left[ U_k(\phi) - U_k(0)\right] = 
-\frac{1}{8 \pi} \left[
\ln \left( 1 + \frac{U''_k(\phi)}{k^2} \right) 
- \ln \left( 1 + \frac{U''_k(0)}{k^2} \right) \right].
\label{evpot} \ee
For the numerical integration of the above equation we use the
algorithms described in ref.~\cite{num}.
The first step of an iterative solution of eq.~(\ref{evpot}) gives
\cite{iterative}
\be
U_k^{(1)}(\phi)-U_k^{(1)}(0)=
U_{k_0}(\phi)-U_{k_0}(0)
+
\frac{1}{2}\ln\left[\frac{\det[-\partial^2+k^2+U''_{k_1}(\phi)]}{
\det[-\partial^2+k^2_0+U''_{k_1}(\phi)]}
\frac{\det[-\partial^2+k^2_0+U''_{k_1}(0)]}{
\det[-\partial^2+k^2+U''_{k_1}(0)]}\right].
\label{iter} \ee
The scale $k_1$ in the above expression can be chosen 
arbitrarily anywhere between $k_0$ and $k$, as the induced
uncertainty for $U_k^{(1)}(\phi)$ corresponds to a higher-order 
contribution in the iterative procedure.
For $k_1=k_0$, $k\to 0$, eq.~(\ref{iter}) is a regularized one-loop
approximation to the effective
potential. Due to the ratio of determinants, only
momentum modes with $k^2\lta q^2 \lta k_0^2$ are effectively included
in the momentum integrals.
The above expression demonstrates the form of ultraviolet regularization
of fluctuation determinants that is consistent with the cutoff 
procedure that leads to the evolution equation for the potential. An analogous
regularization must be used for the fluctuation
determinants in the expression for the nucleation rate.

\subsection{The nucleation rate}
The calculation of the nucleation rate proceeds in complete analogy
to the one in ref.~\cite{first} which we outlined in the introduction.
We define the theory through the potential at a scale $k_0$ below the
temperature, so that the behaviour is effectively two-dimensional.
We then integrate the evolution equation down to a scale 
$k \gta k_f$, where we compute the bubble-nucleation rate. 
In practice, $k_f^2$ is taken $10\%$ larger than the absolute value of
the curvature at the top of the barrier.
The potential has two minima:
the stable (true)
one located at $\phi=\phi_t$, and the unstable (false) one at
$\phi=\phi_f=0$.
The nucleation rate is exponentially suppressed by 
the action $S_k$ (the free energy rescaled by the temperature) 
of the saddle-point configuration
$\phibounce(r)$ that is associated with tunnelling. 
This is an $SO(2)$-symmetric 
solution of
the classical equations of motion 
which interpolates between 
the local maxima of the potential
$-U_k(\phi)$. It satisfies the equation 
\be
{d^2\phibounce\over dr^2}+\frac{1}{r}~{d\phibounce\over dr}=
U'_k(\phibounce),
\label{eom} \ee
with the boundary conditions 
$\phibounce\rightarrow 0$ for $r \rightarrow \infty$ and
$d\phibounce/dr= 0$ for $r =0$.
The action $S_k$ of the saddle point is given by 
\be
S_k=2 \pi
\int_0^\infty 
\left[ \frac{1}{2}
\left(\frac{d\phibounce(r)}{dr}\right)^2
+U_k(\phibounce(r)) -U_k(0) \right]
r\,dr\equiv S_k^t+S_k^v.
\label{action} \ee 
The profile of the saddle point can be easily computed 
with the ``shooting'' method~\cite{shooting}. 
A consistency check for our solution is provided by the fact 
that $S_k^v=0$ for two-dimensional theories.

The bubble-nucleation rate is determined through eq.~(\ref{rate0})
in terms of the potential $U_k(\phi)$. The explicit temperature dependence
is absorbed in the definition of effective two-dimensional parameters
according to eq.~(\ref{fivethree}).
An appropriate regularization is implemented in order to control the 
ultraviolet divergence of the prefactor. 
The type of regularization 
is dictated by the one-loop effective potential,
given by eq.~(\ref{iter}) in our scheme.
This equation indicates that fluctuation
determinants computed within the low-energy theory must be 
replaced by appropriate ratios of determinants.
We emphasize that the matching of the regularization scheme for the
prefactor with the cutoff used for the derivation of the 
evolution equation for the potential~(\ref{evpot}) is crucial for the 
consistency of our method. 
Finally, the nucleation rate in two dimensions is given by 
\beq
I&=&A_{k} \exp({-S_k})
\nonumber \\
\riga{where}\\[-2mm]
A_{k}&=& \frac{E_0}{2\pi} \frac{S_k}{2\pi}
\left|
\frac{\det'\left[-\partial^2+U''_k(\phibounce(r)) \right]}
{\det \left[ -\partial^2+k^2 +U''_k(\phibounce(r))\right]}
~\frac{\det\left[-\partial^2+k^2+U''_k(0) \right]}
{\det\left[-\partial^2+U''_k(0)\right]}
\right|^{-1/2}.
\label{rrate} \eeq

The differential operators that appear in eq.~(\ref{rrate}) 
have the general form
\begin{eqnsystem}{sys:W}
\Op_{\kappa\alpha}&=&-\partial^2 +m_{\kappa}^2+\alpha W_{k}(r) 
\label{op} \\
\riga{where}\\[-2mm]
m_{\kappa}^2&\equiv&U''_k(0)+\kappa k^2,\\
W_{k}(r)&\equiv&U''_k(\phibounce(r))-U''_k(0),\label{ak}
\end{eqnsystem}
with $\kx,\alpha=0$ or 1.
As the $\Op_{\kappa\alpha}$ operators are $SO(2)$-symmetric, 
it is convenient to use polar coordinates and express their 
eigenfunctions as $\psi_n(r,\varphi)=e^{in\varphi} u_n(r)/\sqrt{r}$.
This leads to 
\beq
\det \Op_{\kappa\alpha}
&=&\prod_{n=-\infty}^\infty \det \Op_{n\kappa\alpha}
\nonumber \\
\Op_{n\kappa\alpha}&=&-\frac{d^2}{dr^2}
+\frac{n^2-\frac{1}{4}}{r^2}+m_{\kappa}^2+\alpha W_{k}(r).
\label{opl} \eeq

The computation of such determinants is made possible by 
a theorem~\cite{erice} that relates ratios
of determinants to solutions of ordinary differential equations.
In particular, we have  
\be
g_{n\kappa}\equiv
\frac{\det \Op_{n \kappa 1}}{\det \Op_{n \kappa 0}}=
\frac{y_{n\kappa1}(r\to \infty)}{y_{n\kappa0}(r \to \infty)},
\label{theorem} \ee
where $y_{n\kappa\alpha}(r)$ is the solution of the differential equation
\be
\left[-\frac{d^2}{dr^2}
+\frac{n^2-\frac{1}{4}}{r^2}
+m_{\kappa}^2+\alpha W_{k}(r)\right]y_{n\kappa\alpha}(r)=0,
\label{diffeq} \ee
with the behaviour
$y_{n\kappa\alpha}(r)\propto r^{|n|+\frac{1}{2}}$ for $r\to 0$.
For example, $y_{n\kappa0}$ are proportional to modified Bessel functions:
$y_{n\kappa0}\propto \sqrt{r}~I_{|n|}(m_\kappa r).$
Equations such as~(\ref{diffeq}) can be solved numerically with 
Mathematica~\cite{Mathematica}. 
Since opposite values of $n$ lead to identical determinants,
the final expression for the nucleation rate can be written as  
\beq
I &=& \frac{1}{2 \pi}
\left(\frac{S_k}{2\pi}\right)\exp\left(-S_k\right)
\prod_{n=0}^\infty c_{n},
\nonumber \\
c_{0} &=& \left( \frac{E_0^2 g_{01}}{\left| g_{00} \right|}
\right)^{1/2},~~~~~~~~~~
c_{1} = {g_{11}\over g'_{10}},~~~~~~~~~~
c_{n} = {g_{n 1}\over g_{n 0}}.
\label{fin} \eeq
The calculation of $c_{1}$ is slightly complicated because of the necessity to
eliminate the zero eigenvalue in $g'_{10}$.~(The two zero eigenvalues of 
the operator $-\partial^2+U''_k(\phibounce(r))$ are included in the 
equal factors $g_{-10}$ and $g_{10}$).  
Also the (unique) negative eigenvalue $-E_0^2$ 
must be computed for the determination of $c_{0}$. 
How these steps are achieved is described in ref.~\cite{first}. 
For sufficiently large $n$, one can compute $c_n$ analytically using
first-order perturbation theory in $W_k(r)$ \cite{previous,first}.
We find 
\beq
g_{n\kappa} \approx 1+\frac{1}{2n} \int r\,W_k(r)\,dr,\qquad
c_n \approx 1-\frac{1}{4n^3}k^2\int r^3 \, W_k(r)\,dr.
\label{apppr} \eeq
These expressions are very useful for the evaluation of the prefactor, as
only $c_n$ for small values of $n$ need to be computed numerically.

\setcounter{equation}{0}
\renewcommand{\theequation}{\thesection.\arabic{equation}}

\begin{figure}[t]
\begin{center}\hspace{-5mm}
\begin{picture}(17,9)
\putps(0,-0.4)(-2.1,-0.5){figEx}{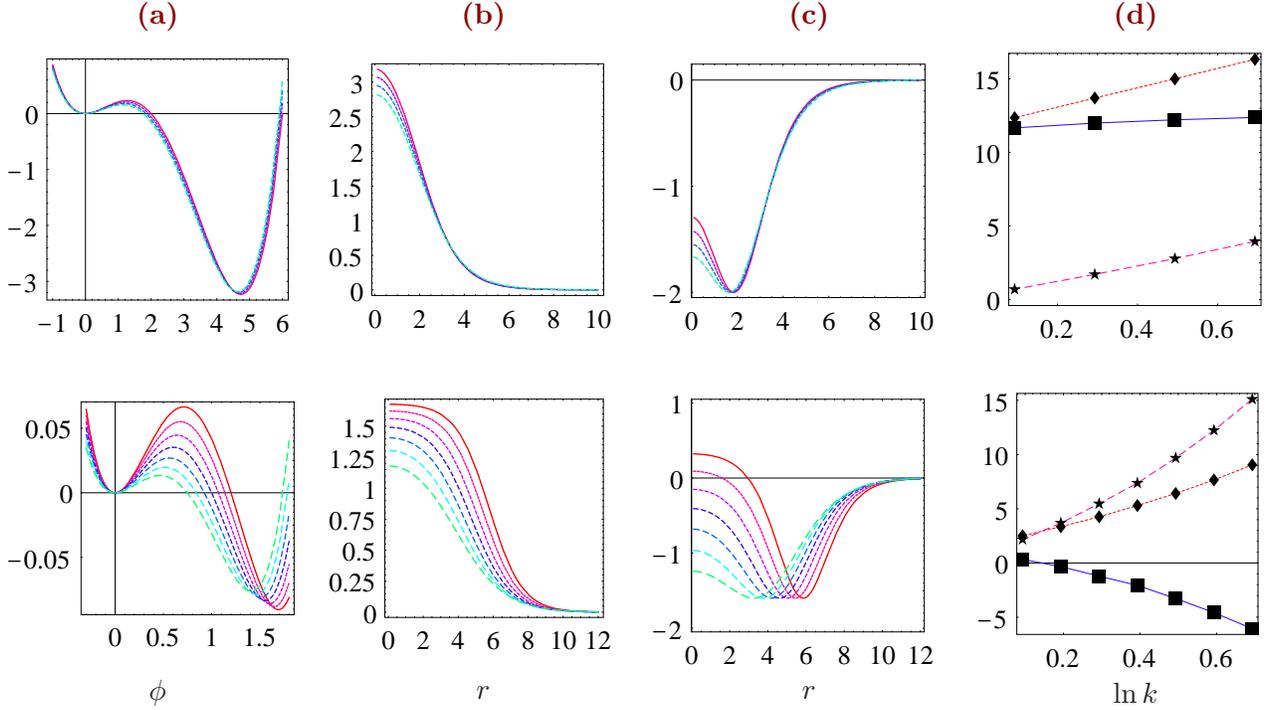}\Red
\put(2.5,8.5){\makebox(0,0){{\bf (a)}}}
\put(6.83,8.5){\makebox(0,0){{\bf (b)}}}
\put(11.16,8.5){\makebox(0,0){{\bf (c)}}}
\put(15.5,8.5){\makebox(0,0){{\bf (d)}}}
\Black
\put(2.5,-0.5){\makebox(0,0){$\phi$}}
\put(6.83,-0.5){\makebox(0,0){$r$}}
\put(11.16,-0.5){\makebox(0,0){$r$}}
\put(15.5,-0.5){\makebox(0,0){$\ln k$}}
\end{picture}
\vspace{1cm}
\caption[SP]{\em The steps in the computation of the nucleation rate:
{\rm (a)} Potentials $V_k(\phi)$;
{\rm (b)} Saddle points $\phi_b(r)$;
{\rm (c)} $W_k(r)$, given by eq.~(\ref{ak});
{\rm (d)} Results for the saddle-point action 
$S_k$ (diamonds), prefactor $\ln\left(A_k/m^3\right)$ (stars) and nucleation
rate $-\ln \left(I/m^3\right)$ (squares).
The first row corresponds to a model with a potential $U_{k_0}(\phi)$ given
by eq.~(\ref{eq:two20}) with $\gamma/m^2=-2$,
$h/m^2=1/3$, while 
the second one to a model with $\gamma/m^2=-4.5$,
$h/m^2=2$.
All dimensionful quantities are given in units of $m$.
\label{fig:Ex}}
\end{center}\end{figure}

\section{A sample computation}
We are interested in potentials that have the approximate form 
\be
V(\phi) =
\frac{m^2}{2} \phi^2
+\frac{\gamma}{6} \phi^3
+\frac{h}{8}  \phi^4.
\label{eq:two20} \ee
The explicit
temperature dependence has been absorbed in $\gamma$,
$h$ and $\phi$ according to eqs.~(\ref{fivethree}). 
Through a shift $\phi\to\phi+c$ the cubic term in eq.~(\ref{eq:two20})
can be eliminated 
in favour of a term linear in $\phi$.
The resulting potential
describes a statistical system of the Ising universality class
in the presence of an external magnetic field.
We assume that the potential of eq.~(\ref{eq:two20}) describes the
theory at some initial scale $k_0$ not far below the temperature,
similarly to refs.~\cite{first}--\cite{third}. Its form
at lower scales can be determined by integrating the evolution equation
(\ref{evpot}) numerically, or by using the approximate
solution of eq.~(\ref{iter}).

The various steps in our calculation are summarized in fig.~1,
for two models described by a potential $U_{k_0}(\phi)$ given
by eq.~(\ref{eq:two20}).
In the following we express all dimensionful quantities in terms of the
arbitrary mass scale $m$.
In the first row we present results for a theory 
with
$\gamma/m^2=-2$,
$h/m^2=1/3$. 
In (a) 
we present the evolution of the potential $U_k(\phi)$ as the scale
$k$ is lowered.
We always shift the metastable vacuum to
$\phi=0$. 
The solid line corresponds to $k_0/m=2$, while the
line with longest dashes (that has the smallest barrier height)
corresponds to $k_f/m=1.1$. At the scale $k_f$ the negative 
curvature at the top of the barrier is slightly larger than 
$-k_f^2$. 
This is the point in the evolution of the potential
where configurations that 
interpolate between the minima start becoming relevant
in the functional integral that defines the coarse-grained potential
\cite{convex}. 
For this reason,
we stop the evolution at this point. 
We observe that, for low  $k$, the absolute minimum of the potential
settles at a non-zero value of $\phi$. A significant barrier
separates it from the metastable minimum at $\phi=0$.
The profile of the saddle point $\phibounce(r)$
is plotted in (b) in units of $m$
for the same sequence of scales.  
We observe a 
variation of the value of the field $\phi$ in the center of the critical 
bubble for different $k$.
This is reflected in the form of the quantity $W_k(r)$, defined in
eq.~(\ref{ak}), which we plot in~(c).

Our results for the nucleation rate are presented in (d).
On the horizontal axis we give the values of $\ln(k/m)$.
The dark diamonds correspond to the values of the action $S_k$ 
of the saddle point at the scale $k$. 
The stars indicate the values of 
$\ln ( A_k/m^3 )$. 
We observe a logarithmic $k$ dependence 
for both these quantities.
The value of $A_k$ is expected to decrease for smaller $k$, because
$k$ acts as the effective ultraviolet cutoff in the calculation 
of the fluctuation determinants in $A_k$. For smaller $k$,
fewer fluctuations with wavelengths above an increasing length scale
$\sim 1/k$ contribute explicitly to the fluctuation determinants. 
The logarithmic dependence on $k$ is the reflection of the logarithmic
ultraviolet divergence of the unregularized prefactor in two dimensions. 
The dark squares in (d) give our results for 
$-\ln(I/m^3) 
= S_k-\ln ( A_k/m^3 )$.
The logarithmic $k$ dependence largely cancels between 
$S_k$ and $\ln ( A_k/m^3 )$, so that 
$\ln(I/m^3)$ is almost constant.
The small residual dependence on $k$ can be used to estimate the 
contribution of the next order in the expansion around the saddle point.
This contribution is expected to be smaller than
$\ln ( A_k/m^3 )$. 
This behaviour confirms that the
nucleation rate should be independent of the scale $k$ that 
we introduced as a calculational tool. 

In the second row we present the calculation of the nucleation rate for 
a model with a larger coupling $h/m^2=2$, and 
$\gamma/m^2=-4.5$,
$k_0/m=2$, $k_f/m=1.1$. 
We observe a more pronounced $k$ dependence of the potential, saddle-point
profile and function $W_k(r)$.
The most important aspect of the comparison of the two models
concerns the relative values of 
$S_k$ and $\ln ( A_k/m^3 )$.
For the first model,
the contribution of the prefactor to the
nucleation rate is much smaller than that of the action of the 
saddle point. The main role of the prefactor is 
to remove the logarithmic $k$ dependence from $I/m^3$. 
For the second model,
$S_k$ and $\ln ( A_k/m^3 )$ are comparable.
This indicates that the effects of fluctuations
are enhanced.
Moreover, the prefactor fails to cancel the $k$ dependence
of the saddle-point action. 
The reason is that 
the next order in the expansion around the saddle point is 
important and can no longer be neglected.
This establishes the limit of validity of 
homogeneous nucleation theory~\cite{langer} for two-dimensional
systems, in agreement with the studies of refs.~\cite{first}--\cite{third}
in three dimensions.

\section{Comparison with lattice studies}
\subsection{Matching the lattice theory}
A main objective of this work is the comparison of our results
with the lattice study of ref.~\cite{twopone}.
This requires a precise definition of the form of the potential.
We must make sure that we consider a theory identical to 
the one simulated on the lattice.
As we cannot match exactly the bare parameters of the lattice action, it
is more convenient to guarantee that the low energy renormalized theory
is the same in our model and ref.~\cite{twopone}.
In the latter work,
through a redefinition of the field and the distance, the
one-loop renormalized action (free energy rescaled with respect to the
temperature) 
of the (2+1)-dimensional theory at high temperature 
is expressed as
\be
S=
\frac{1}{\theta} \int d^2 \xt \biggl\{
- \frac{1}{2} \left( 1-\frac{\theta}{48\pi}\right) 
\phit \, \tilde{\partial}^2 \phit 
+\Vt(\phit) + \frac{\theta}{8\pi}\left[
\Vt''(\phit)-\Vt''(\phit)\ln \left(\Vt''(\phit)\right)
\right] \biggr\},
\label{renorm} \ee
with
\be
\Vt(\phit)= \frac{1}{2}\phit^2 - \frac{1}{6}\phit^3+ \frac{1}{24} \lx \phit^4. 
\label{pot} \ee
For $\lx<0$ the potential becomes unbounded from below, while for
$\lx=1/3$ it has two equivalent minima. For this reason, we
consider $\lx$ values in the interval $(0,1/3)$.
Higher-loop corrections are expected to be proportional to larger powers
of $\theta/8\pi$. This implies that 
the theory that is simulated on the lattice corresponds to a renormalized
action in the continuum given by eqs.~(\ref{renorm}), (\ref{pot}) only if 
$\theta$ is not much larger than 1. In the opposite case, the renormalized
action cannot be determined perturbatively.

In our approach the theory is defined at some initial scale $k_0$ and the
integration of evolution equations such as eq.~(\ref{evpot}) generates 
the low-energy structure. For $k=0$ the effective average action of 
eq.~(\ref{twoeleven}) must be identified with the action of eq.~(\ref{renorm}).
We neglect wavefunction renormalization effects,
which can be seen from eq.~(\ref{renorm})
to be a good approximation for values of $\theta$ not much larger than 1.
The form of the potential at some non-zero scale 
$k$ can be inferred by 
demanding that the iterative solution of eq.~(\ref{iter}) reproduces the 
potential of eq.~(\ref{renorm}) for $k=0$. 
In particular, by choosing $k=k_1=0$ in
eq.~(\ref{iter}) and renaming $k_0$ as $k$ we find 
\be
U_{k}(\phi) \approx V(\phi) + \frac{1}{8\pi} V''(\phi)
- \frac{1}{8\pi} \left(k^2+V''(\phi) \right)
\ln \left(\frac{k^2+V''(\phi)}{m^2}
\right),
\label{incond} \ee
where $V(\phi)$ is given by eq.~(\ref{eq:two20}) with 
\beq
\frac{\gamma}{m^2}=-\sqrt{\theta},\qquad
\frac{h}{m^2}=\frac{1}{3}\, \theta\, \lx.
\label{param} \eeq
In eq.~(\ref{incond}) we have neglected 
terms $\sim(8 \pi)^{-2}$ and
terms $\sim (8 \pi)^{-1}$
in the arguments of the logarithms. 
It is clear from the above that the dimensionless coupling that controls the
validity of the perturbative expansion is $\theta \lx/3$. For this reason,
perturbation theory is expected to break down for $\theta \gta 3 /\lx$. 

In summary, eq.~(\ref{incond}) is an approximate solution of the evolution
equation (\ref{evpot}) (at the first level of an iterative procedure)
which is consistent with 
eq.~(\ref{renorm}) that determines the renormalized action of 
ref.~\cite{twopone}. 
The matching between the renormalized and the lattice actions
is accurate only at the one-loop level
(the region of validity of eq.~(\ref{renorm})).
The approximate 
solution of eq.~(\ref{incond}), which results at the first level of the 
iterative procedure, has a similar 
region of applicability.
Morever, this approximation 
is not valid for values of
$k$ that render negative the argument of the logarithm in the 
right-hand side of eq.~(\ref{incond}), i.e. for
$k^2 < {\rm max} \left\{-V''(\phi)\right\}$. 
This implies that
eq.~(\ref{renorm}) is trustable only in the convex regions of
the potential (for which $\Vt''(\phi) > 0$)
and should not
be expected to lead to reliable predictions for the nucleation rate. This 
is in agreement with the conclusion of 
ref.~\cite{twopone} that 
the determination of the bubble-nucleation rate 
through the real part of 
eq.~(\ref{renorm}) does not lead to consistency 
with the lattice results. On the other hand, 
eq.~(\ref{renorm}) is perfectly valid in the convex regions of the
potential, where it can be matched with 
eq.~(\ref{incond}) for $k = 0$.
In the following, for our predictions of the bubble-nucleation rates and
the comparison with the lattice results, we
rely on the approximate solution of eq.~(\ref{incond}),
instead of integrating eq.~(\ref{evpot}) numerically. 
The numerical solution, which is more accurate than eq.~(\ref{incond}),
does not offer increased 
precision in our comparison with the lattice data, as the determination of
the renormalized theory is valid only at the 
one-loop level\footnote{We also perform checks of the corrections arising from 
integrating the full evolution equation
(\ref{evpot}). These corrections are very small in
the parameter region for which the expansion around the saddle point
is convergent.}.

\subsection{Comparison with the lattice results}
In fig.~2 we present a comparison of results obtained through our method
with the lattice results of fig.~1 of ref.~\cite{twopone}. 
For each of several values of $\lx$ we vary the parameter $\theta$
and determine the couplings $\gamma$, $h$ according to eqs.~(\ref{param}).
The coarse-grained potential is then given by eq.~(\ref{incond}) for 
$k \geq k_f$.
The diamonds denote the saddle-point action $S_k$. 
For every choice of $\lx$, $\theta$ we determine $S_k$
at two scales: $1.2\,k_f$ and $2\,k_f$. The light-grey region between
the corresponding points gives an indication of the
$k$ dependence $S_k$.
The bubble-nucleation rate $-\ln\left(I/m^3\right)$ is denoted 
by dark squares. 
The dark-grey region between the values obtained at 
$1.2\,k_f$ and $2\,k_f$ gives a good check of the convergence of the
expansion around the saddle-point. If this region is thin, the
prefactor is in general small and
cancels the $k$ dependence of the action.
The dark circles denote the results for the nucleation rate from
the lattice study of ref.~\cite{twopone}. The dashed straight lines
correspond to the action of
the saddle-point computed from the `tree-level'
potential of eq.~(\ref{eq:two20}).

\begin{figure}[t]
\begin{center}\hspace{-5mm}
\begin{picture}(17,11)
\putps(0,-0.4)(0,-0.5){figRes}{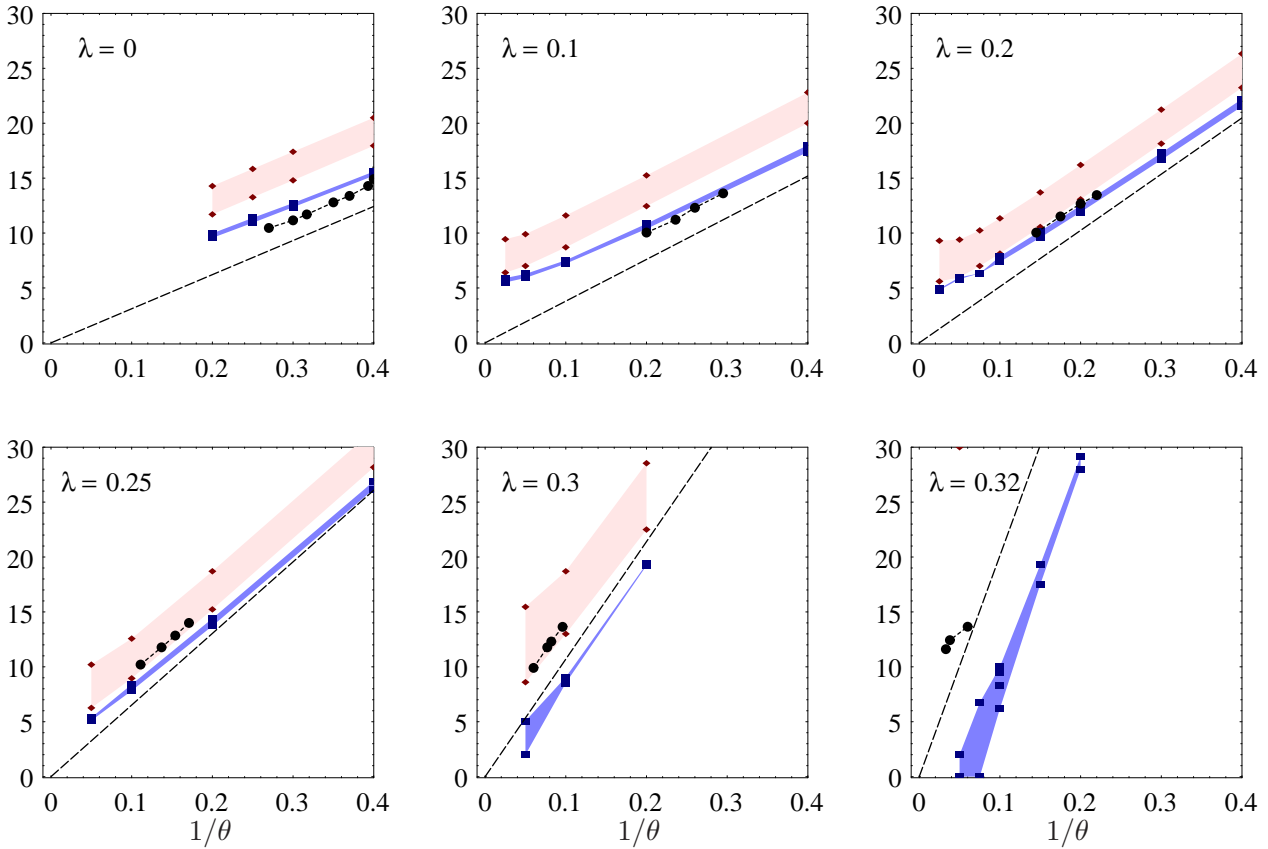}
\put(3.2,-0.5){\makebox(0,0){$1/\theta$}}
\put(8.9,-0.5){\makebox(0,0){$1/\theta$}}
\put(14.6,-0.5){\makebox(0,0){$1/\theta$}}
\end{picture}
\vspace{1cm}
\caption[SP]{\em Comparison of our method with lattice studies:
Diamonds denote the saddle-point action $S_k$ 
and squares the bubble-nucleation rate 
$-\ln\left(I/m^3\right)$ 
for $k=1.2\,k_f$ and $2\,k_f$. 
Dark circles denote the results for the nucleation rate from
the lattice study of ref.~\cite{twopone}. 
Finally, the dashed straight lines
correspond to the action of 
the saddle-point computed from the 
potential of eq.~(\ref{eq:two20}).
\label{fig:Res}}
\end{center}\end{figure}

For $\lx=0$
the potential of eq.~(\ref{eq:two20}) is unbounded from below and 
the procedure we outlined in the previous paragraphs for matching
the theory simulated on the lattice  may be problematic. However, one
may consider these results as applying to 
the limit $\lx\to 0$, so that no conceptual problems arise. 
The values of $-\ln(I/m^3)$ computed at $1.2\,k_f$ and $2\,k_f$
are equal to a very good approximation, which
confirms the convergence of the expansion around the saddle-point and 
the reliability of the calculation. The $k$ dependence of
the saddle-point action is cancelled by the prefactor, so that the 
total nucleation rate is $k$ independent. Moreover, the
prefactor is always significantly smaller than the
saddle-point action.  
The circles indicate the results of
the lattice simulations of ref.~\cite{twopone}. 
The agreement with the lattice predictions is 
good. More specifically, it is clear that the contribution of
the prefactor is crucial for the correct determination of the
total bubble-nucleation rate. Similar conclusions can be drawn 
for $\lx=0.1$ and $\lx=0.2$.

For larger values of $\lx$ the lattice simulations have been performed
only for $\theta$ significantly larger than 1. For smaller
$\theta$, nucleation events become too rare to be observable 
on the lattice. As we discussed
earlier, the matching between the lattice and the renormalized actions
becomes imprecise for large $\theta$. 
This indicates that we should expect
deviations of our results from the lattice ones, as the theory of
eq.~(\ref{renorm}) may be different than the simulated one. 
These deviations start becoming apparent for the value $\lx=0.25$, for
which the lattice simulations were performed with $\theta \sim 10$--20.
For $\lx\geq 0.3$ the lattice results are in a region in which 
the internal consistency criteria of our method for the reliability of
the expansion around the saddle-point are not  
satisfied any more.

The consistency of our calculation is achieved  
for $1/\theta \gta 0.12$ even for $\lx=0.32$.
However, 
the breakdown of the expansion around the saddle point is apparent
for $1/\theta \lta 0.12$.
The $k$ dependence of the predicted bubble-nucleation rate is 
strong\footnote{The additional squares for $1/\theta=0.1$ correspond
to results from the numerical integration of eq.~(\ref{evpot}).}.
The prefactor becomes comparable to the saddle-point
action and the higher-order corrections are expected to be large.
The $k$ dependence of $S_k$ is very large. For this reason we 
have not given values of $S_k$ in this case.
This behaviour
indicates that the field fluctuations become significant.
Typically, these fluctuations enhance the total rate and for
$1/\theta \lta 0.08$ can even compensate 
the exponential suppression. Several similar examples 
were given in refs.~\cite{first}--\cite{third}.  

The reason for this behaviour can be traced to the form of the differential
operators in the prefactor (see eqs. 
eqs.~(\ref{rrate})--(\ref{ak})).
This prefactor, before regularization, involves the ratio 
$\det'(-\partial^2+U''_k(0)+W_{k}(r))/\det(-\partial^2+U''_k(0))$, with
$W_{k}(r)=U''_k\left(\phi_b(r)\right)
-U''_k(0)$. The function $W_{k}(r)$ always has a minimum away from
$r=0$ (see figs.~1c and 1g), 
where it takes negative values. As a result the lowest
eigenvalues of the operator $\det'(-\partial^2+U''_k(0)+W_{k}(r))$ are smaller
than those of $\det(-\partial^2+U''_k(0))$. The elimination of
the very large eigenvalues from the determinants
through regularization does not affect this
fact and the prefactor $A_{k}$ is always larger than 1. Moreover,
in cases such as those depicted in fig.~2f it becomes exponentially large
because of the proliferation of low eigenvalues in 
$\det'(-\partial^2+U''_k(0)+W_{k}(r))$.
In physical terms, this implies the 
existence of a large class of field configurations of free energy comparable
to that of the saddle-point. Despite the fact that they are not 
saddle points of the free energy 
(they are rather deformations of such a point)
and are, therefore, unstable, they result in a dramatic increase of
the nucleation rate. This picture is similar to that of
``subcritical bubbles'' of ref.~\cite{gleiser}. 
In ref.~\cite{nonpert}
the nucleation rate was
computed by first calculating a corrected potential that incorporates 
the effect of such non-perturbative configurations. The pre-exponential factor
must be assumed to be of order 1 in this approach, as the effect of most
deformations of the critical bubble has already been taken into account in
the potential.
In our approach the non-perturbative effects are incorporated
through the prefactor.
Both methods lead to similar conclusions for the enhancement of 
the total nucleation rate.

A final comment concerns the $\theta$ dependence of our results for 
the nucleation rate in fig.~2. For a given value of $\lx$, we observe a 
linear dependence of $-\ln(I/m^3)$ on $1/\theta$
in the regions where the calculation is consistent. This is explained by the
fact that the dominant $\theta$ dependence of the potential
arises from the first term in the right-hand side of eq.~(\ref{incond}).
This is in agreement with the findings of ref.~\cite{twopone}. 
In the latter 
work, this linear dependence was not observed for $\lx=0.3$ and 0.32.
The reason is that the range of $\theta$ used in the simulations was so large
that the first term in the right-hand side of eq.~(\ref{incond}) ceased to
be dominant. Our results for small $\theta$ display the expected behaviour
even for $\lx=0.32$.

\setcounter{equation}{0}
\renewcommand{\theequation}{\thesection.\arabic{equation}}

\section{Conclusions}

In this paper we applied our approach for the calculation of
bubble-nucleation rates to (2+1)-dimen\-sional theories at non-zero
temperature. We studied these theories at coarse-graining scales $k$
below the temperature, where they display an effective two-dimensional
behaviour.
This provided the opportunity to check several points of our approach
that depend on the dimensionality of the system. For example,
the evolution equation for the coarse-grained potential
has a different form than the one in the effective three-dimensional
systems we studied in the past. Also, the
pre-exponential factor in the bubble-nucleation rate
has a logarithmic ultraviolet divergence before regularization,
instead of the leading linear divergence in three dimensions.
This divergence is reflected in the leading logarithmic dependence of
the regularized prefactor on the scale $k$
that acts as an ultraviolet cutoff (fig.~1).
More crucially, the complementarity between the $k$ dependence of
the prefactor and the saddle-point action was observed again. 
This is a crucial point that guarantees that a physical quantity, such
as the bubble-nucleation rate, is independent of the scale $k$ that 
we introduced as a calculational tool. 
The validity of Langer's theory of homogeneous nucleation
was confirmed, as long as the prefactor gave a contribution to
the nucleation rate smaller than the leading exponential
suppression by the action of the saddle point. 

Another important aspect of this work concerns the comparison
of our results with data from lattice simulations. This constitutes
a stringent quantitative test of our method. We found good agreement
between our results and the lattice data when 
the renormalized action for the 
theory that is simulated on the lattice is known (fig.~2). In
these cases, we first match the renormalized action and then
compute the nucleation rate.
For part of the range of the lattice parameters
this is not possible, because the renormalized action 
cannot be obtained from the lattice action
perturbatively. However,
the internal consistency criteria of our method
provide a test of the reliability of our results in all cases.

\paragraph{Acknowledgements} We  would like to thank
M. Gleiser and C. Wetterich for helpful discussions.
The work of N.T. was supported by the E.C. under TMR contract 
No. ERBFMRX--CT96--0090.

\end{document}